\documentclass[journal]{IEEEtran}
\usepackage{amsmath,amssymb}
\usepackage[mathscr]{eucal}
\usepackage{cite,verbatim,color}
\usepackage{epsfig,multicol}

\newcommand{\CC}{{\mathcal C}_{fb,nc}^{MAC}}
\newcommand{\Ri}{{\mathcal R}_{i}^{MAC}}
\newcommand{\Ro}{{\mathcal R}_{o}^{MAC}}
\DeclareMathOperator{\Exp}{{\mathbb E}} 
\newcommand{\abs}[1]{\lvert#1\rvert}
\newtheorem{thm}{Theorem}[section]

\newtheorem{defn}{Definition}[section]

\newtheorem{rem}{Remark}[section]

\begin{document}

\title{On the Capacity of Multiple Access Channels with State Information and Feedback}

\author{
Wei Wu, Sriram Vishwanath and Ari Arapostathis
\thanks{
This research was supported 
in part by NSF grants CCF-0448181,CCF-0552741, ECS-0218207 and ECS-0225448, THECB ARP 010115-0013-2006, the Office of Naval Research through the Electric Ship Research and Development Consortium and a grant from Freescale Semiconductor Corporation. Wei Wu was also supported by the Hemphill-Gilmore Student Endowed Fellowship through the University of Texas at Austin. 
}
\thanks{The authors are with Wireless Networking and Communications Group, 
Department of Electrical and Computer Engineering, The University of
Texas at Austin, Austin, TX 78712, USA (e-mail: \{wwu,sriram,ari\}@ece.utexas.edu).}
}
\maketitle

\begin{abstract}
In this paper, the multiple access channel (MAC) with channel state is analyzed in a scenario where a) the channel state is known non-causally to the transmitters and b) there is perfect causal feedback from the receiver to the transmitters. An achievable region and an outer bound are found for a discrete memoryless MAC that extend existing results, bringing together ideas from the two separate domains of MAC with state and MAC with feedback. Although this achievable region does not match the outer bound in general, special cases where they meet are identified. 

In the case of a Gaussian MAC, a specialized achievable region is found by using a combination of dirty paper coding and a generalization of the Schalkwijk-Kailath \cite{SK:66part1}, Ozarow \cite{Ozarow:84MAC} and Merhav-Weissman \cite{Merhav:ISIT05fb} schemes, and this region is found to be  capacity achieving. Specifically, it is shown that additive Gaussian interference that is known non-causally to the transmitter causes no loss in capacity for the Gaussian MAC with feedback.
 
\end{abstract}


\section{Introduction}
The capacity of channels with perfect feedback \cite{Kramer:IT02feedback,Sahai:ISIT05Boost,WVA:Allerton05, Younghan:Allerton05} is of great interest, as it provides us with an outer limit on the performance of any feedback-based scheme. 
Although the capacity of single user memoryless channels is unaffected by feedback \cite{Shannon:56feedback}, that of multiple access channels (MAC) is known to be enhanced by feedback from the receiver to both transmitters \cite{Ozarow:84MAC}. 
In the two-user Gaussian MAC case, the entire capacity region can be found in closed form employing Ozarow's ingenious extension \cite{Ozarow:84MAC} of the Schalkwijk-Kailath (SK) coding scheme \cite{SK:66part1} that enables limited cooperation between transmitters by using the feedback information.

On a parallel track, the capacity of channels with state has been studied under a varied set of assumptions on state knowledge \cite{Shannon:58causal, Gelfand:80noncausal,Gamal:83noncausal}. 
An important result in this family is when the state in a discrete memoryless (DM) single user channel is known non-causally to the transmitter \cite{Gelfand:80noncausal,Gamal:83noncausal}. 
In the all-Gaussian (additive noise and interference) case, this result, combined with a clever choice for the auxiliary variable translates into the Costa-coding result \cite{Costa:IT83DPC}. 
The capacity of discrete memoryless multiple-access channels is still an open problem, but in the Gaussian case a result that is similar the Costa result is shown in \cite{Younghan:ISIT04MDPC}.

Recently, these two classes of problems were combined in the study of memoryless single user channels with non-causal state knowledge and feedback \cite{Merhav:ISIT05fb}. 
It was shown in this work that there is no capacity gain from feedback, and in the all-Gaussian case, interference causes no loss in capacity.

\subsection{Our contributions} 
In this work, from one point of view, we are interested in understanding the impact of feedback in a MAC channel with state, and from the other, the impact of state on a MAC channel with feedback. As a concrete research challenge, we focus on a MAC channel with state and feedback where the state variable is known non-causally to both the transmitters, and find a general achievable region and outer bound on the capacity region. We find this achievable region and outer bound to meet for a non-trivial class of channels which includes the binary symmetric and binary erasure channels.
 
Next, we specialize our study to the case of a Gaussian MAC with additive interference and feedback. 
The achievable region we obtain for this scenario coincides with a simple outer bound on the system, and thus results in the capacity region. 
>From this capacity characterization, we find that:
\begin{itemize}
\item Feedback enhances the capacity of the MAC channel with state.
\item Interference when non-causally known at the transmitter has no impact on the capacity region of the Gaussian 2 user MAC channel.
\end{itemize}
 Thus, our results are analogous to the Merhav-Weissman \cite{Merhav:ISIT05fb} and Costa \cite{Costa:IT83DPC} results for the single-user case.

\subsection{Organization}
The rest of the paper is organized as follows. In Section ~\ref{sec:def}, basic definitions and notation used are introduced. The main results of the paper for the discrete memoryless channel case are presented in Section ~\ref{sec:results}. The Gaussian case is handled in Section \ref{sec:Gaussian}. Detailed proofs for Sections \ref{sec:results}  are presented in Section \ref{sec:proof} and the correspondence concludes with Section{sec:conclude}. 

\section{Notations and Preliminaries}\label{sec:def}
\subsection{Notation}
We adopt the following notation throughout the correspondence. 
For matrix $A$, $A^T$, $A^{-1}$ denote the transpose and inverse of $A$ respectively.
Random variables (RVs) will be noted by capital letters, while their realizations will be denoted by the respective lower case letters. 
$X_m^n$ denotes the random vector $(X_m, \ldots, X_n)$, and $X_i^n$ denotes the random vector $(X_{i,1}, \ldots, X_{i,n})$. 
Both $X_{i, j}$ and $X_i(j)$ is used to denote the $j$-th random variable of a random vector $X_i$. 
$\Exp{[X]}$ denotes the expectation of random variable $X$ and the correlation coefficient of two scaler random variable $X_1$, $X_2$ is defined as 
\begin{equation*}
\rho_{X_1 X_2}=\frac{\Exp[X_1 X_2]}{\Exp[X_1 X_1^T] \Exp{[X_2 X_2^T]}}\,.
\end{equation*}
The alphabet of a random variable $X$ will be designated by a calligraphic letter $\mathcal{X}$, and that of the $n$-fold Cartesian power of $\mathcal{X}$ will be denoted as $\mathcal{X}^n$.
\begin{equation*}
X \Rightarrow Y \Rightarrow Z
\end{equation*}
will be used to denote the conditional independence of $X$ and $Z$ given $Y$.

\subsection{Models and definitions}
A two-user \emph{multiple access channel with random parameters} $(\mathcal{X}_1, \mathcal{X}_2, \mathcal{S}, \mathcal{Y}, P(y|x_1, x_2, s))$ is a channel with two input alphabets $\mathcal{X}_1$ $\mathcal{X}_2$; state space $\mathcal{S}$, output alphabet $\mathcal{Y}$, and transition probability $P(y|x_1, x_2, s)$. 
The states $s$ take values in $\mathcal{S}$ according to the probability mass function (PMF) $P(s)$. It is assumed that the channel and the state are both memoryless, namely,
\begin{equation*}
P(y^n|x_1^n, x_2^n, s^n)=\prod_{i=1}^{n} P(y_i|x_{1,i}, x_{2,i}, s_i)
\end{equation*}   
and 
\begin{equation*}
P(s^n)=\prod_{i=1}^{n} P(s_i)\,.
\end{equation*}

Here we assume the state variable is noncausally known. 
Both \emph{noncausal} state information and \emph{feedback} are incorporated into the channel model via the definition of a noncausal feedback code $(R_1, R_2, n)$ as follows. 
\begin{defn}
An $(R_1, R_2, n)$ code for the MAC $(\mathcal{X}_1, \mathcal{X}_2, \mathcal{S}, \mathcal{Y}, P(y|x_1, x_2, s))$ with feedback and noncausal state information is defined by encoding functions and decoding functions:
\begin{enumerate}
\item The encoding functions for user $i$ are the mappings 
\begin{multline*}
f_{i,k}: \{1, \ldots, 2^{n R_i}\} \times \mathcal{S}^n \times \mathcal{Y}^{k-1} \to \mathcal{X}_i\,, \\
i=1,2,\, k=1,2, \ldots, n \,,
\end{multline*}
or in other words, for the message of user $i$, $w_i \in \{1, 2, \ldots, 2^{n R_i}\}$, $i=1,2$, the channel input is expressed as 
\begin{equation}
x_{i,k}=f_{i,k}(w_i, y^{k-1}, s^n)\,,
\end{equation}
where $s^n$ is the noncausal state information of the whole block and $y^{k-1}$ is the perfect feedback of channel output up to time $t-1$.
\item The decoding functions for the receiver are the mappings 
\begin{equation*}
g: \mathcal{Y}^n \to \{1, \ldots, 2^{n R_1}\} \times \{1, \ldots, 2^{n R_2}\}\,, 
\end{equation*}
or in other words, the decoder gives the estimates of the two messages $w_1$, $w_2$, $\hat{w}=(\hat{w}_1, \hat{w}_2)^T$,
\begin{equation}
\hat{w}=g(y^n)\,.
\end{equation}
\end{enumerate}
\end{defn}

We shall use the average probability of error criterion $P_e$ assuming that the messages $(w_1, w_2)$ are drawn according to uniform distribution over $\{1, \ldots, 2^{n R_1}\} \times \{1, \ldots, 2^{n R_2}\}$. 

\begin{defn}
A rate pair $(R_1, R_2)$ per channel use is achievable for the MAC with feedback and noncausal state information if there exists a sequence of $(R_1, R_2, n)$ codes such that $P_e \to 0$ as $n \to \infty$. 
\end{defn}

The capacity region of MAC with feedback and noncausal state information, $\CC$, is the closure of the set of all achievable rates. $\CC$ is known to be convex by time multiplexing of achievable rates.

\section{Main results} \label{sec:results}
Let $\mathcal{P}$ stand for the collection of all RVs $(S, U, X_1, X_2, V_1, V_2, Y)$ while $U$, $V_1$, $V_2$ are auxiliary variables introduced to form a Markov chain
\begin{equation*}
S \Rightarrow U \Rightarrow ((X_1, V_1), (X_2, V_2)) \Rightarrow Y \,.
\end{equation*}
Define $\Ri$ to be the set of all rate pairs $(R_1, R_2)$ such that 
\begin{eqnarray}
R_1 &\le& I(X_1;Y|X_2,U,S) \nonumber \\
R_2 &\le& I(X_2;Y|X_1,U,S) \label{eq:achievability}\\
R_1 + R_2 &\le& I(V_1,V_2;Y) - I(V_1,V_2;S) \nonumber
\end{eqnarray}
with joint distribution $P(u|s)P(v_1,x_1|u,s)P(v_2,x_2|u,s)$. 
As stated in the next theorem, the set $\Ri$ is an inner bound on the capacity region $\CC$.
\begin{thm} \label{thm:inner}
The capacity region of the MAC channel $(\mathcal{X}_1, \mathcal{X}_2, \mathcal{S}, \mathcal{Y}, P(y|x_1, x_2, s))$, with feedback and noncausal state information at both encoders, satisfies 
\begin{equation*}
\Ri \subseteq \CC\,.
\end{equation*}
\end{thm}

In this achievable region, the auxiliary variable ``U'' reflects the amount of common information shared between the two transmitters, while ``$V_i$'' is the auxiliary variable associated with the message from Transmitter $i$. This expression is highly intuitive - the sum rate expression in \eqref{eq:achievability} resembles a generalized Gel'fand-Pinsker expression where there is non-causal side information at the transmitters and no information at the receivers; while the two individual constraints reflect the scenario where both the transmitter and the receiver know the channel state.
 
A proof of this is given in Section \ref{sec:inner}. This proof builds on the Cover-Leung \cite{Cover:IT81MACfb} and Gel'fand-Pinsker arguments \cite{Gelfand:80noncausal}. It differs from them in the following ways:

\begin{enumerate}
\renewcommand{\theenumi}{\roman{enumi}}
\item binning is used to determine common transmission ($U$), 
\item backward decoding is employed at the receiver; and
\item sequences $v_1^n$ and $v_2^n$ are not placed in separate bins at each transmitter. 
\end{enumerate}

All three changes are introduced both to facilitate the result and as simplifying mechanisms to make the proof tractable.

The outer bound is stated next. 
Define $\Ro$ to be the set of all rate pairs $(R_1, R_2)$ such that 
\begin{eqnarray}
R_1 &\le& I(V_1;Y|V_2)-I(V_1;S|V_2) \nonumber\\
R_2 &\le& I(V_2;Y|V_1)-I(V_2;S|V_1) \label{eq:outer}\\
R_1+R_2 &\le& I(V_1,V_2;Y)-I(V_1;V_2;S) \nonumber
\end{eqnarray}
for all joint distribution $P(v_1,v_2,x_1,x_2|s)$. 

\begin{thm}\label{thm:outer}
The capacity region of the MAC channel $(\mathcal{X}_1, \mathcal{X}_2, \mathcal{S}, \mathcal{Y}, P(y|x_1, x_2, s))$, with feedback and noncausal state information at both encoders, satisfies 
\begin{equation*}
\CC \subseteq \Ro\,.
\end{equation*}
\end{thm}

This outer bound expression is quiet intuitive. As $V_1$,$V_2$ represent the messages from  Transmitters 1 and 2, the region resembles a combination of the multiple access capacity region combined with a generalization of the Gel'fand-Pinsker expression. In that spirit, the proof for this outer bound is an extension of the Gel'fand-Pinsker arguments \cite{Gelfand:80noncausal} with one important modification.  For the proof and the modification to the Gel'fand-Pinsker argument, see Section ~\ref{sec:outer}.with one important modification.  For the proof and the modification to the Gel'fand-Pinsker argument, see Section ~\ref{sec:outer}.

Although the achievable region and outer bounds do not meet in general, one can determine non-trivial sufficient conditions for them to do so. 
Consider the class of MAC channels (called class $\Gamma$) that satisfy either  
\begin{equation}\label{eq:DMC_cond}
H(X_1|S, X_2, Y)=0 \quad \text{or} \quad H(X_2|S,X_1, Y)=0\,. 
\end{equation}	 

\begin{thm}\label{thm:DMC_cap}
The capacity region of a MAC channel in class $\Gamma$, $(\mathcal{X}_1, \mathcal{X}_2, \mathcal{S}, \mathcal{Y}, P(y|x_1, x_2, s))$, with feedback and noncausal state information at both encoders, is given by
\begin{equation*}
\CC = \Ri\,.
\end{equation*}
\end{thm}

The proof of Theorem ~\ref{thm:DMC_cap} is provided in Section ~\ref{sec:DMC_cap}. 
The key idea is to use the condition in \eqref{eq:DMC_cond} to prove a tighter outer bound  for the class channels belonging to $\Gamma$. 

The condition in \eqref{eq:DMC_cond} covers a wide class of discrete memoryless MAC channels. 
For the input sets and state set $\mathcal{X}_1=\mathcal{X}_2=\mathcal{S}=\{0,1\}$, both the binary MAC adder channel, 
\begin{equation*} 
Y=(X_1+X_2+S)\, \text{mod}\, 2 \,,\quad \mathcal{Y}=\{0,1\}\,,
\end{equation*}
and the MAC ``erasure-type'' channel,
\begin{equation*}
Y=(X_1+X_2+S)\, \text{mod}\, 3 \,,\quad \mathcal{Y}=\{0,1,2\}\,,
\end{equation*}
satisfy \eqref{eq:DMC_cond}, thus the capacity region can be characterized using Theorem ~\ref{thm:DMC_cap}.

On the other hand, the Gaussian MAC channel does not belong to class $\Gamma$ thus Theorem ~\ref{thm:DMC_cap} does not apply. 
In the next section, we will develop a coding strategy tailored to the Gaussian MAC channel.
This coding scheme builds on Ozarow's feedback coding scheme \cite{Ozarow:84MAC} and the dirty paper coding strategy by Costa \cite{Costa:IT83DPC} to obtain the full capacity region. 

\section{The capacity region of the Gaussian MAC channel with random parameters}
\label{sec:Gaussian}

Consider the Gaussian multiple access channel
\begin{equation*}
Y(k) = X_1(k) + X_2(k) + S(k) + Z(k)\,, 
\end{equation*} 
where $\{S(k)\}$ denotes the interfering signal, which are i.i.d. Gaussian random variables with $\Exp S(k)=0$ and $\sigma^2_{S}=\Exp S^2(k) \leq \infty$. It is assumed to be known to the encoder non-causally. $Z(k)$ in this model is zero-mean, white Gaussian noise with variance $\sigma^2_{Z}$. 
The average powers for the two transmitters are assumed to be $P_1$, $P_2$ respectively. 

One of the first results for this class of channels is in \cite{Merhav:ISIT05fb}, where Merhav and Weissman show feedback does not increase the capacity of the point-to-point version of this channel, but one can significantly reduce the coding complexity and achieve \emph{doubly exponential} error exponent by using an extension of the Schalkwijk-Kailath (SK) \cite{SK:66part1} coding scheme. 

In this section, we will show for the Gaussian MAC channel with feedback and noncausal state information at the transmitters, the capacity is as if the interference were absent, with a double exponential decay in probability of error.

The coding scheme we employ to show our results is a mixture of two different schemes - the Ozarow coding scheme \cite{Ozarow:84MAC} and the Merhav-Weissman scheme \cite{Merhav:ISIT05fb}; each one of which is a generalization of the original Schalkwijk-Kailath coding scheme.

\subsection{Coding Scheme that achieves sum capacity}
\label{sec:sum}

 We use the following notation in this subsection: $k$ is the time index, $i$ is the transmitter index (denoted by $T_i$). The main idea of this scheme is to set an initial condition, and over time send corrections (innovations) to the receiver to allow the receiver to converge to this initial condition. Here is the coding scheme:

{\bf Before transmission}. Given a message for transmitter $T_i$, $m_i$, $m_i \in \{0, 1, \ldots, M_i-1\}$, $M_i=2^{n R_i}$, map $m_i$ to a point on the real line as follows: $\theta^0_i=(m_i+1/2)/M_i$. Define $a_i \triangleq (-1)^{i-1} 2^{R_i}$. 
	Given the noncausal state $S^n=(S(1), \ldots, S(n))$, compute a {\it precancelling} message $\theta_i(k)$ for $k=2, \ldots, n$, as
	\begin{equation*}
	\theta_i(p)=\theta_i^0+\sum_{j=p+1}^n \frac{l_i S(p)}{a_i^{p-2}}\,, 
	\end{equation*}  
	where $p \in \{2,\ldots,n\}$ and  $l_i$ is a scaling factor that will be presented later in this section.

{\bf Initialization}. This is what we call the first two transmissions $(k=1,2)$. At time $k=1$, $T_2$ sends nothing while $T_1$ sends
	$$X_1(1)=\theta_1(2)-S(1)\,.$$
	The receiver obtains $Y_1=X_1(1)+S(1)+Z(1)=\theta_1(2)+Z(1)$ and receiver finds an initial estimate ($\hat{\theta}_1$) for $\theta_1^0$ to be $\hat{\theta}_1(1)=Y(1)$.  At time $k=2$, $T_1$ sends nothing while $T_2$ sends 
	$$X_2(2)=\theta_2(2)-S(2)\,.$$
	Then receiver sets its initial estimate for $\theta_2^0$ to be 
	$$\hat{\theta}_2(2)=Y(2)=\theta_2(2)+Z(2)\,.$$

{\bf Estimation recursion}. This is the remainder of the transmissions $k=3,\ldots,n$. Defining $\epsilon(k) \triangleq \hat{\theta}_i(k)-\theta_i(k)$. 
	At time $k$,  $T_i$ transmits a scaled version of $\epsilon_k$: 
	\begin{equation}
	X_i(k)=a_i^{k-2} \epsilon_i(k-1).
	\end{equation}

At the end of time $k$,	the receiver updates its estimate of message $\theta_i^0$ as:
	\begin{equation}\label{eq:receiver}
\hat{\theta}_i(k)=\hat{\theta}_i(k-1)-a_i^{-(k-2)} l_i Y(k)\,,
	\end{equation}

 while Transmitter $T_i$ updates $\epsilon_i$ as
	\begin{align*}
	\epsilon_i(k) &=\hat{\theta}_i(k)-\theta_i(k) \\
	&= \bigl(\hat{\theta}_i(k-1)-\theta_i(k-1)\bigr)+ \frac{l_i S(k)}{a_i^{k-2}} \\
	& \quad\quad\quad- \frac{l_i (X_1(k)+X_2(k)+S(k)+Z(k))}{a_i^{k-2}} \\
	&= \epsilon_i(k-1)- \frac{l_i (X_1(k)+X_2(k)+Z(k))}{a_i^{k-2}}\,.
	\end{align*}

{\bf Analysis}

	The estimation error at the receiver for $T_i$ at the end of the $kth$ transmission, denoted $\tilde{\epsilon}_i(k)$ is  

\begin{equation}
\tilde{\epsilon}_i(k) =\hat{\theta}(k)-\theta_i^0 =\epsilon_i(k)+(\theta_i(k)-\theta_i^0)\,. \nonumber
\end{equation}

	At the end of the block $k=n$, $\theta_i(n)=\theta_i^0$ thus $\tilde{\epsilon}_i(n)=\epsilon_i(n)$. 
	As long as $\epsilon_i(n)$ goes to zero as $n \to \infty$, the estimation error at the receiver goes to zero also.
	
We denote $Y'=X_1+X_2+Z$. Notice that all the steps taken above were {\it linear}. We can combine them into one system equation as:

	\begin{equation} \label{eq:recursion}
	X(k+1)= A X(k)- LY'(k)\,,
	\end{equation}
	where 
	\begin{equation*}
	X =[X_1, X_2]^T,\quad  A ={\text{diag}}(a_1, a_2), \quad L=[a_1 l_1, a_2 l_2]^T.
	\end{equation*}
	Along the lines of \cite{Ozarow:84MAC} and \cite{WVA:Allerton05}, we chose the optimal $L$ that minimizes the mean squared error as: 
\begin{equation*}
L(k)=\frac{\Exp[Y'(k) X(k)]}{\Exp[Y'(k)^2]}\,.
\end{equation*}
	Denoting $Q=\Exp[X(k) X^T(k)]$, we obtain
	\begin{equation}
	Q(k+1)=A\bigl[Q(k)- Q(k) H^T (H Q(k) H^T+ \sigma_Z^2)^{-1}H Q(k)\bigr]A\,, \nonumber
	\end{equation}
	from (\ref{eq:recursion}) where $H=[1 \quad 1]$ and $Y'=H X + Z$.
	Since $(H,A)$ is detectable, the matrix recursion above converges and $Q(k) \to Q$ as $k$ goes to $\infty$, satisfying 
	\begin{equation}\label{eq:dare}
	Q=A\bigl[Q- Q H^T (H Q H^T+ \sigma_Z^2)^{-1}H Q\bigr]A\,.
	\end{equation}
	From (\ref{eq:dare}), we solve for $Q$ in terms of $a_i$ and $\sigma_z^2$. Note that the diagonal elements of $Q$ are the individual power constraints $P_i$, and the off-diagonal element is the correlation between the transmit signals $\rho$. These turn out to be:
\eqref{eq:dare},
	\begin{align}
	P_1 & =\frac{(a_1^2-1)(\abs{a_1 a_2}+1)^2}{(\abs{a_1}+\abs{a_2})^2} \sigma_Z^2 \label{eq:MAC_P1} \\
	P_2&=\frac{(a_2^2-1)(\abs{a_1 a_2}+1)^2}{(\abs{a_1}+\abs{a_2})^2}\sigma_Z^2\,, \label{eq:MAC_P2}
	\end{align}
	and the correlation coefficient $\rho$ between $X_1$ and $X_2$ satisfies,
	\begin{equation} \label{eq:MAC_rho}
\rho=\sqrt{\frac{(a_1^2-1)(a_2^2-1)}{(\abs{a_1}\abs{a_2}+1)^2}}\,. 
	\end{equation}
	One can find the achievable rate $(R_1, R_2)$ in terms of the power constraints $(P_1, P_2)$ and $\rho$ by rewriting the equations above as:
	\begin{equation}\label{eq:sum_cap}
	\begin{split}
	R_1 &= \frac{1}{2} \log\Bigl(1+\frac{P_1 (1-\rho^2)}{\sigma_Z^2}\Bigr) \\
	R_2 &= \frac{1}{2} \log\Bigl(1+\frac{P_2 (1-\rho^2)}{\sigma^2_Z}\Bigr) \\
	R_1+R_2 & = \frac{1}{2}\log\Bigl(1+\frac{P_1+P_2+2\rho\sqrt{P_1 P_2}}{\sigma^2_Z}\Bigr) \,. 
	\end{split}
	\end{equation}
	
It is well-known that \eqref{eq:sum_cap} is the sum-capacity point for the normal two-user Gaussian MAC with feedback, which is certainly the outer bound of the counterpart with state variable. Thus \eqref{eq:sum_cap} is the also the sum-capacity for Gaussian MAC with feedback and noncausal state. 
	
\begin{rem}
Note that this precancellation strategy requires non-causal knowledge of the interference at both transmitters and causal may not be sufficient. Also note that the receiver updates the estimates \emph{as if} there is no interference and by the end of the block, and due to the pre-cancelation, the estimation errors are not affected by the state.    
\end{rem} 
\begin{rem}
Note $\Exp[X_1(1)^2] < \infty$ and $\Exp[X_2(2)^2] < \infty$, due to the average power constraint, as long as the power assumption at the \emph{initialization stage} is finite, the average power is asymptotically close to $P_i$ as $n \to \infty$.
Moreover, $\{S(k)\}$ is not restricted to be Gaussian  to achieve sum-capacity \eqref{eq:sum_cap}. All that is required is $\sigma_S^2 <\infty$. 
\end{rem}
\begin{rem}
Using the same type of pre-canceling, one can extend the coding scheme to Gaussian broadcast (BC) channel and Gaussian interference channel  \cite{Kramer:IT02feedback} and \cite{WVA:Allerton05}.
\end{rem}

\subsection{Hybrid coding that achieve other points in the capacity region}

In the previous section, we focus on sum-capacity \eqref{eq:sum_cap}.
In this section, we combine the coding scheme in Subsection \ref{sec:sum} with dirty-paper coding (Costa coding) to obtain the full capacity region. The strategy here  mimics the approach used in \cite{Ozarow:84MAC} with an important difference - instead of superposition coding, we employ dirty paper coding.

Let one transmitter, say $T_1$, have two messages which we call $m_1^{(1)}$ and $m_1^{(2)}$ at rates $R_1^{(1)}$ and $R_1^{(2)}$ respectively. 
Given the noncausal state $S^n$, let $T_1$ use power $\alpha P_1$, ($0 \leq \alpha \leq 1$) to transmit $m_1^{(2)}$ and use the remainder $\bar{\alpha} P_1$ ($\bar{\alpha}=1-\alpha$) to transmit $m_1^{(1)}$. $m_1^{(1)}$ is transmitted using the feedback coding scheme described in Subsection \ref{sec:sum}, while $m_1^{(2)}$ is sent independently. Meanwhile, $T_2$ uses all his power to transmit $m_2$ using the feedback coding scheme in Subsection \ref{sec:sum}.

Decoding: at the receiver, messages $m_1^{(1)}$ and $m_2$ are first decoded by treating the code letters of $m_1^{(2)}$ as noise. Finally, $m_1^{(2)}$ is decoded. 

According to \eqref{eq:sum_cap}, $m_1^{(1)}$ and $m_2$ will be transmitted reliably iff 
\begin{align}
R_1^{(1)} & \leq \frac{1}{2} \log\Bigl(1+\frac{\bar{\alpha} P_1 (1-\rho^2)}{\sigma_Z^2+\alpha P_1}\Bigr) \label{eq:R1_1}\\
	R_2 & \leq \frac{1}{2} \log\Bigl(1+\frac{P_2 (1-\rho^2)}{\sigma^2_Z+\alpha P_1}\Bigr) \label{eq:R2} \,,
\end{align}
where $\rho$ satisfies
\begin{multline} \label{eq:eq_cond}
\Bigl(1+\frac{\bar{\alpha} P_1 (1-\rho^2)}{\sigma_Z^2+\alpha P_1}\Bigr) \Bigl(1+\frac{P_2 (1-\rho^2)}{\sigma^2_Z+\alpha P_1}\Bigr) \\
=\Bigl(1+\frac{\bar{\alpha} P_1+P_2+2\rho\sqrt{\bar{\alpha} P_1 P_2}}{\sigma^2_Z+\alpha P_1}\Bigr) \,.
\end{multline}
At the end of the block, after decoding $m_1^{(1)}$ and $m_2$, one can, with high probability, subtract the feedback coding from the channel output and obtain $\tilde{Y}^n$, where 
\begin{equation*}
\tilde{Y}^n=X_1^n(m_1^{(2)})+S^n+Z^n\,,
\end{equation*} 
Here $X_1^n(m_1^{(2)})$ denotes the dirty paper coded transmit vector corresponding to message $m_1^{(2)}$. 
Then $m_1^{(2)}$ can be decoded reliably at rate 
\begin{equation}\label{eq:R1_2}
R_1^{(2)} \leq \frac{1}{2}\log\Bigl(1+\frac{\alpha P_1}{\sigma_Z^2}\Bigr)\,.
\end{equation}
By adding \eqref{eq:R1_1} and \eqref{eq:R1_2}, we obtain the total rate for $T_1$ as:
\begin{align*}
R_1 & = R_1^{(1)} + R_1^{(2)} \nonumber\\
& \leq \frac{1}{2}\log\Bigl[1+ \frac{(1-\bar{\alpha} \rho^2) P_1}{\sigma_Z^2}\Bigr]\,.
\end{align*}
Using \eqref{eq:eq_cond} and \eqref{eq:R2}, we can obtain 
\begin{align*}
R_2 & \leq \frac{1}{2} \log \Bigl[1+\frac{\bar{\alpha} P_1+P_2+2\rho\sqrt{\bar{\alpha} P_1 P_2}}{\sigma^2_Z+\alpha P_1}\Bigr] \\
& - \frac{1}{2} \log \bigl(1+\frac{\bar{\alpha} P_1 (1-\rho^2)}{\sigma_Z^2+\alpha P_1}\bigr) \nonumber \\
& = \frac{1}{2}\log \Bigl[\frac{\sigma_Z^2+P_1+P_2+2\rho\sqrt{\bar{\alpha} P_1 P_2}}{\sigma_Z^2+(1-\bar{\alpha} \rho^2) P_1}\Bigr]
\end{align*}
Note by defining $\rho'=\sqrt{\bar{\alpha}}\rho$, $\abs{\rho'} \leq 1$ and for any $0 \leq \rho' \leq \rho$,
\begin{align*}
R_1 &\leq \frac{1}{2}\log\Bigl[1+ \frac{(1-\rho'^2) P_1}{\sigma_Z^2}\Bigr]\\
R_2 & \leq \frac{1}{2}\log \Bigl[\frac{\sigma_Z^2+P_1+P_2+2\rho'\sqrt{P_1 P_2}}{\sigma_Z^2+(1-\rho'^2) P_1}\Bigr]
\end{align*}
is achievable. 
A similar region can be obtained when $T_2$ splits the power to combine feedback coding and dirty paper coding.
Thus the achievable region can be written as 
\begin{multline}\label{eq:cap_region}
\bigcup_{0 \leq \rho \leq 1}\Bigl\{(R_1, R_2): 0 \leq R_1 \leq \frac{1}{2}\log\big[1+ \frac{(1-\rho^2) P_1}{\sigma_Z^2}\bigr] \\
0 \leq R_2 \leq \frac{1}{2}\log\big[1+ \frac{(1-\rho^2) P_2}{\sigma_Z^2}\bigr] \\
0 \leq R_1+R_2 \leq \frac{1}{2}\log \bigl[1+\frac{P_1+P_2+2\rho\sqrt{P_1 P_2}}{\sigma_Z^2}\bigr] \Bigr\}\,. 
\end{multline}
Note \eqref{eq:cap_region} is the same as the capacity region of Gaussian MAC with feedback but without interference \cite{Ozarow:84MAC}, which is clearly the outer bound of the channel with interference, thus \eqref{eq:cap_region} is also the capacity region of Gaussian MAC with feedback and noncausal interference, $C_{MAC}^{fb, S}$. 

\section{Proofs} \label{sec:proof}
\subsection{Proof of Theorem ~\ref{thm:inner}} \label{sec:inner}
Here we provide a proof for Theorem ~\ref{thm:inner}, 
the achievability of \eqref{eq:achievability}.

\begin{figure*}[htbp]
\centerline{\scalebox{1.0}{\input{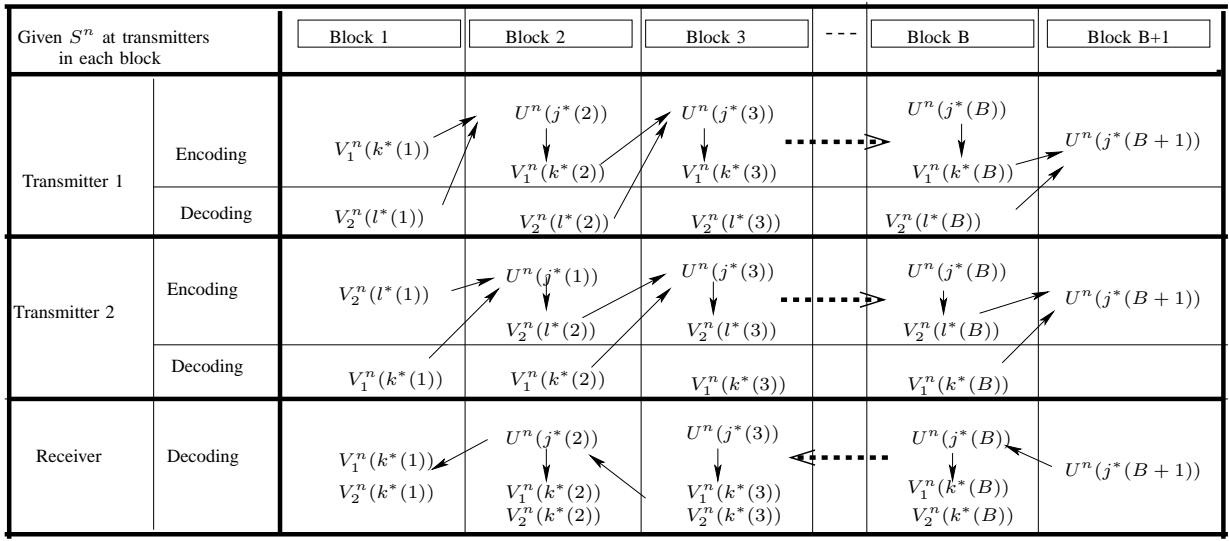}}}
\caption{The Modified Cover-Leung Coding Procedure}
\label{fig:feedback}
\end{figure*}

{\bf Code Generation:}  $2^{nR_0}$ strongly typical sequences $\sim p(u^n|s^n)$ are first generated. At transmitter $i$, for each sequence $u^n(j), j \in \{1, \ldots, 2^{n R_0}\}$, $2^{n R_i}$ $V_i^n$ sequences are generated using $p(v_i^n|u^n,s^n)$ and are indexed using $k \in \{1,\ldots, 2^{nR_1}\}$ and $l \in \{1,\ldots,2^{nR_2}\}$ respectively.

The index pairs $(k,l)$ are thrown uniformly into $2^{nR_0}$ bins such that each bin receives $2^{n(R_1+R_2-R_0)}$ of them. These bins are indexed by $j,  j \in \{1, \ldots, 2^{n R_0}\}$.

A total of $B$ ($B$ large) messages are sent over blocks of length $n$ each.  At the beginning of block $b$, each transmitter successfully decoders the other transmitter's message sent in block $b-1$, while the receiver waits till the end of transmission to decode all messages. Using shared information from block $b-1$, the transmitters cooperatively transmit ``cloud centers'' $u^n(j)$  in block $b$. This process is diagrammatically illustrated in Figure~\ref{fig:feedback}.

{\bf Encoding}: Suppose $j^*(b) \in \{1, \ldots, 2^{n R_0}\}$ is the common index to be sent by the two transmitters in block $b$. Let $k^*(b)$ be the message to be transmitted by Transmitter 1 and $l^*(b)$ that by Transmitter 2. Transmitter 1 locates $v_1^n(k^*(b))$ given the particular $u^n(j^*(b))$ and $s^n(b)$, then determines a sequence $x_1^n$ that is jointly typical with the pair $v_1^n,s^n$ and transmits it.

Similarly, Transmitter 2 locates $v_2^n(l^*(b))$ that is jointly typical with $u^n(k^*(b)),s^n(b)$, then generates a sequence $x_2^n$ that is jointly typical with this particular $v_2^n$ and $s^n$. $x_1^n$ and $x_2^n$ are transmitted in block $b$.

{\bf Decoding:} We employ backward decoding at the receiver, i.e., we wait till all $B+1$ transmissions are complete before we begin decoding. Each transmitter, however, decodes the other transmitter's message at the end of each block.

Decoding at Transmitter 1:  Transmitter 1 looks for a unique index $l(b)$ such that the set $(x_1^n(k^*(b)),v_2^n(l),u^n(j^*(b)),s^n(b),y^n(b))$ are jointly typical. Note here that $x_1^n(k^*(b))$,$u^n(j^*(b))$,$s^n(b)$ are all known at transmitter 1, and so $l(b)$ can be determined uniquely iff
\begin{equation}
R_2 < I(X_2;Y|X_1,U,S)  
\end{equation}

Now, $j_1^*(b+1)$ at Transmitter 1 is determined as the bin index in which the pair $k^*(b),l(b)$ lie.

Decoding at Transmitter 2: Similarly, Transmitter 2 can determine the unique bin index corresponding to Transmitter 1's message if $R_1 < I(X_1;Y|X_2,U,S)$.
 $j_2^*(b+1)$ is determined as the bin index to which the pair  $k^*(b),l(b)$ belong. 

For $n$ large, with high probability we have  $j_1^*(b+1) = j_2^*(b+1) \triangleq j^*(b+1)$.

Backward decoding at receiver:  In block $B+1$ only the common ``cloud center'' $u^n(j^*(B+1))$ is communicated at a rate of $R_0 < I(U;Y)$. Using $j^*(B+1)$ as the bin index, the subset of all possible choices for $k(B),l(B)$ are determined. We call this subset $S_B$. Note that the cardinality of $S_B$ is $2^{n(R_1+R_2-R_0 \pm \epsilon)}$.

In block $B$, the cloud center $u^n(j(B))$ is first determined, which can be performed at a rate $R_0 < I(U;Y)$. Next, the unique pair of indices $k(B),l(B)$ in the set $S_B$ are located, if they exist, such that $(u^n(j(B)),v_1^n(k(B)),v_2^n(l(B)),y^n)$ are all jointly typical.

The set of possible errors determines the bound on the pair $R_1,R_2$. Here we ignore trivial error cases and concentrate on those that provide bounds on rates. 
Specifically, the event 
\begin{multline*} E_{k'l'}=\{(u^n(j(B)),v_1^n(k'),v_2^n(l'),y^n) \in T_{\epsilon}^n | \\
(v_1^n(k'),v_2^n(l'),s^n) \in T_{\epsilon}^n\}\}
\end{multline*} 
is of concern when either $k' \ne k^*(B)$ or $l' \ne l^*(B)$.

So
\begin{eqnarray*}
P_e &=& P(\bigcup_{k' \ne k^*(B), l' \ne l^*(B), (k',l') \in S(B)} E_{k'l'}) \nonumber\\
    &\le& \sum_{i=2}^{2^n(R_1+R_2-R_0)} P(E_{k'l'}) \nonumber \\
    &\le& 2^{n(R_1+R_2-R_0)} 2^{-n(I(V_1,V_2;Y|U)-I(V_1,V_2;S)- 6 \epsilon)} \label{eqn:bound}
\end{eqnarray*}

Thus it is sufficient if \\ $R_1+R_2-R_0 \le I(V_1,V_2;Y|U)-I(V_1,V_2;S)$.

Proof of \eqref{eqn:bound}. By Bayes' rule
\begin{eqnarray*}
P(E_{k'l'})&= \frac{P(\{(u^n(j(B)),v_1^n(k'),v_2^n(l'),y^n) \in T_{\epsilon}^n , (v_1^n(k'),v_2^n(l'),s^n) \in T_{\epsilon}^n\})}{P(\{(v_1^n(k'),v_2^n(l'),s^n) \in T_{\epsilon}^n\})}\\
&\le 2^{-n(I(V_1,V_2;Y|U)-3\epsilon)}/2^{-n(I(V_1;V_2;S)+3 \epsilon)}
\end{eqnarray*}

Thus we have the result.

\subsection{Proof of the outer bound of \eqref{eq:outer}} \label{sec:outer}

The outer bound given in \eqref{eq:outer} is proved here. 
\begin{align}
n R_1 &= H(W_1) \leq I(W_1; Y^n) \label{eq:one}\\
&  \leq I(W_1; Y^n | W_2) \label{eq:two}\\
& = \sum_{i} I(W_1; Y_i | W_2, Y^{i-1}) \nonumber \\
& = \sum_{i} I(W_1, S_{i+1}^n; Y_i| W_2, Y^{i-1}) \nonumber\\
& - \sum_{i} I(S_{i+1}^n; Y_i| W_2, Y^{i-1}, W_1) \label{eq:three}\\
& = \sum_{i} \bigl( I(W_1,Y_i|W_2, Y^{i-1}, S_{i+1}^n)  \nonumber\\ 
& \quad \quad + I(S_{i+1}^n; Y_i|W_2, Y^{i-1}) \label{eq:four}\\
& \quad\quad - I(S_{i+1}^n; Y_i| W_2, Y^{i-1}, W_1)\bigr) \nonumber
\end{align}
Here, \eqref{eq:one} results from Fano's inequality ($n \epsilon_n$ is dropped as a convenience) \cite{ElementsInfoTheory}, \eqref{eq:two} from the fact that conditioning reduces entropy,  \eqref{eq:three}) and \eqref{eq:four} from the chain rule \cite{ElementsInfoTheory}.

Note by the property of mutual information in \cite{Csiszar_Korner} and the fact that $S_i$ is independent of $W_1$, $W_2$ and $S_{i+1}^n$, we have
\begin{align*}
&\sum_{i} I(S_{i+1}^n; Y_i| W_2, Y^{i-1}) \\
&= \sum_{i} I(S_i; Y^{i-1}|W_2, S_{i+1}^n) \\
&=\sum_{i} I(W_2, S_{i+1}^n, Y^{i-1}; S_i)\\
& \sum_{i} I(S_{i+1}^n; Y_i| W_1,W_2, Y^{i-1}) \\
&= \sum_{i} I(S_i;Y^{i-1}|W_1, W_2, S_{i+1}^n) \\
&= \sum_{i} I(W_1, W_2, S_{i+1}^n, Y^{i-1}; S_i)\,. 
\end{align*}
By defining the auxiliary variables $V_1$, $V_2$ as
\begin{equation}\label{eq:aux_V12}
\begin{split}
V_{1,i} & = (W_1, Y^{i-1}, S_{i+1}^n) \\
V_{2,i} & = (W_2, Y^{i-1}, S_{i+1}^n)\,,
\end{split}
\end{equation}
we have that 
\begin{align*}
n R_1 &\leq \sum_{i} \bigl(I(W_1, Y_i | W_2, Y^{i-1}, S_{i+1}^n) + I(W_2, S_{i+1}^n, Y^{i-1}; S_i) \\
& \quad\quad - I(W_1, W_2, S_{i+1}^n, Y^{i-1}; S_i) \bigr) \\
& = \sum_{i} \bigl( I(V_{1,i};Y_i | V_{2,i}) - I(W_1; S_i | Y^{i-1}, S_{i+1}^n, W_2) \bigr) \\
& = \sum_{i} \bigl( I(V_{1,i}; Y_i|V_{2,i}) -  I(V_{1,i}; S_i | V_{2,i}) \bigr)\,.
\end{align*}
In the same way, one can obtain the inequality for $R_2$. For the sum rate, we follow the Gel'fand-Pinsker arguments \cite{Gelfand:80noncausal} almost exactly:
\begin{align*}
n (R_1+R_2) & = H(W_1, W_2) \\
& \leq I(W_1, W_2; Y^n) \\
& \leq \sum_{i} I(W_1, W_2, Y^{i-1}; Y_i) \\
& = \sum_{i} \bigr(I(W_1, W_2, Y^{i-1}, S_{i+1}^n; Y_i) \\
&\quad\quad\quad -I(S_{i+1}^n; Y_i| Y^{i-1}, W_1, W_2)\bigr) \\
& = \sum_{i} \bigr(I(V_{1,i}, V_{2,i}; Y_i) \\
& -I(Y^{i-1}; S_i| W_1, W_2, S_{i+1}^n)\bigr) \\
& = \sum_{i} \bigr(I(V_{1,i}, V_{2,i}; Y_i)-I(V_{1,i}, V_{2,i}; S_i)\bigr) \,.
\end{align*} 

%

\subsection{Proof of Theorem ~\ref{thm:DMC_cap}} \label{sec:DMC_cap}
The achievability is the same as the proof of the general achievable region in Theorem~\ref{thm:inner}. 

Now we prove \eqref{eq:achievability} is also the outer bound.
Without any loss of generality, we may assume $H(X_1|X_2, S, Y)=0$, or there exists  deterministic functions $F_k$, such that,
\begin{equation}\label{eq:X1_fun}
X_{1,k}=F_k (X_{2,k}, S^n, Y^k)\,.
\end{equation}
Then for $R_1$, we have  
\begin{align}
n R_1 & =  H(W_1)=H(W_1|W_2, S^n) \nonumber\\
& \leq I(W_1; Y^n|W_2, S^n) \label{eq:cap_R1_fano}\\
& = \sum_{k=1}^{n}I(W_1;Y_k|W_2, S^n, Y^{k-1}) \nonumber\\
& =\sum_{k=1}^{n} \bigl(H(Y_k|W_2, S^n, Y^{k-1}, X_1^{k-1}, X_2^k) \nonumber\\
& \quad\quad\quad-H(Y_k|W_1, W_2, S^n, Y^{k-1}, X_1^k, X_2^{k})\bigr) \label{eq:cap_R1_fun}\\
&\leq  \sum_{k=1}^{n} \bigl(H(Y_k|S^n,Y^{k-1}, X_1^{k-1}, X_{2,k}) \nonumber \\
& \quad\quad\quad -H(Y_k|X_1^{k-1}, Y^{k-1}, S^n, X_{1,k}, X_{2,k})\bigr) \label{eq:cap_R1_dep} \\
&= \sum_{k=1}^{n}I(X_{1,k};Y_{k}| S^n, Y^{k-1}, X_1^{k-1}, X_{2,k}) \nonumber \\
&= \sum_{k=1}^{n} I(X_{1,k}; Y_{k}| S_k, X_{2,k}, U_k)\,, \nonumber
\end{align} 
where the auxiliary variables $U_k$ are defined as 
\begin{equation*}
U_k=(Y^{k-1},X_1^{k-1}, S^{k-1}, S_{k+1}^{n})\,.
\end{equation*}
Thus for $R_2$, 
\begin{align}
n R_2 & = H(W_2) = H(W_2|W_1, S^n) \nonumber\\
& \leq I(W_2; Y^n|W_1, S^n) \label{eq:cap_R2_fano}\\
& = \sum_{k=1}^{n} I(W_2; Y_k| W_1, S^n, Y^{k-1}) \nonumber\\
& = \sum_{k=1}^{n} \big(H(Y_k|W_1, S^n, Y^{k-1}, X_1^k) \nonumber\\ 
& \quad\quad\quad -H(Y_k|W_1,W_2,S^n,Y^{k-1}, X_1^k, X_2^k)\bigr) \label{eq:cap_R2_fun}\\
& \leq \sum_{k=1}^{n}\bigl(H(Y_k|S^{n},Y^{k-1},X_1^{k-1}, X_{1,k}) \nonumber\\
& \quad\quad\quad - H(Y_k|S^n, Y^{k-1}, X_{1,k}, X_{2,k})\bigr) \label{eq:cap_R2_dep}\\
& =\sum_{k=1}^{n} I(Y_k; X_{2,k}| U_k, S_k, X_{1,k})\,. \nonumber
\end{align}
Note \eqref{eq:cap_R1_fano} and \eqref{eq:cap_R2_fano} are due to Fano's inequality \cite{ElementsInfoTheory}. 
\eqref{eq:cap_R1_fun} and \eqref{eq:cap_R2_fun} are from \eqref{eq:X1_fun} and the fact that the channel input $X_i=G_i(W_i, S^n, Y)$ for feedback coding.   
\eqref{eq:cap_R1_dep} and \eqref{eq:cap_R2_dep} are because 
\begin{equation*}
(W_1, W_2, Y^{k-1}, X_1^{k-1}, X_2^{k-1}) \Rightarrow (X_{1,k}, X_{2,k}, S_k) \Rightarrow Y_k
\end{equation*}
forms a Markov chain.
Now define the auxiliary variables $V_{1,k}$, $V_{2,k}$ as \eqref{eq:aux_V12} and note given $U_k$, $V_{1,k}$ and $V_{2,k}$ are independent, thus the proof for the sum-rate in theorem~\ref{thm:outer} can be applied directly to establish \eqref{eq:achievability}.  

\section{Conclusions}
\label{sec:conclude}

In this paper, both the discrete memoryless and the Gaussian two-user multiple access channel with state and feedback are analyzed, where the state is non-causally known at the transmitters. 
For the discrete memoryless case, both an outer bound and an achievable region are derived, and sufficient conditions under which they meet are obtained. For the all-Gaussian case, the entire capacity region can be found. 
This capacity region is the same as that of a Gaussian MAC with feedback with channel state known to the transmitters and the receiver. 
The proofs in this paper are obtained as generalizations of the Merhav-Weissman, Ozarow, Costa, Gel'fand-Pinsker and Cover-Leung coding schemes.

\section*{Acknowledgment}
The authors would like to thank Dr. Gerhard Kramer for helpful discussions and comments.

\bibliographystyle{IEEEtran}
\bibliography{wei-info}
\end{document}